\documentclass[aps,prl,reprint,onecolumn,superscriptaddress]{revtex4-2} 
\usepackage{ulem}
\usepackage{amsfonts}
\usepackage{amssymb}
\usepackage{amsthm,amsmath}
\usepackage{mathrsfs}
\usepackage{indentfirst}
\usepackage{multirow}
\usepackage{graphicx}
\usepackage{amsmath}
\usepackage[amssymb]{SIunits}
\usepackage{amssymb}
\usepackage{color}
\usepackage[usenames,dvipsnames]{xcolor}
\usepackage{natbib}
\usepackage[T1]{fontenc}
\usepackage[colorlinks=true, citecolor=blue, linkcolor=blue, urlcolor=blue, bookmarks=true]{hyperref}
\usepackage{bm}
\usepackage{booktabs}
\usepackage{multirow}
\usepackage{setspace}

\begin{document}

\title{Freezing and ice aging dynamics in saline water under natural convection}

\author{Feng Wang}
\thanks{Equally contributed authors}
\affiliation{New Cornerstone Science Laboratory, Center for Combustion Energy, Key Laboratory for Thermal Science and Power Engineering of MoE, Department of Energy and Power Engineering, Tsinghua University, China}

\author{Yihong Du}
\thanks{Equally contributed authors}
\affiliation{New Cornerstone Science Laboratory, Center for Combustion Energy, Key Laboratory for Thermal Science and Power Engineering of MoE, Department of Energy and Power Engineering, Tsinghua University, China}

\author{Xueyi Xie}
\affiliation{Department of Engineering Mechanics, School of Aerospace Engineering, Tsinghua University, China}

\author{Enrico Calzavarini}
\affiliation{Univ.\ Lille, Unit\'e de M\'ecanique de Lille - J. Boussinesq (UML) ULR 7512, F-59000 Lille, France}

\author{Chao Sun}
\thanks{chaosun@tsinghua.edu.cn}
\affiliation{New Cornerstone Science Laboratory, Center for Combustion Energy, Key Laboratory for Thermal Science and Power Engineering of MoE, Department of Energy and Power Engineering, Tsinghua University, China}
\affiliation{Department of Engineering Mechanics, School of Aerospace Engineering, Tsinghua University, China}

\date{\today}

\begin{abstract} 
Understanding the coupled dynamics of liquid-solid phase change and fluid flows is crucial in a wide range of geophysical and industrial applications.
When freezing occurs in saline water, the newly formed ice is mushy, with a porous structure that traps the brine within the ice.
In this work, which combines experiments and theoretical analyses, we investigate the long-term evolution of saline ice, comprehensively accounting for the coupled dynamics of multiscale fluid flow, heat and mass transfer, and phase change.
We show that in a closed convective system the rapid formation of a mushy ice layer is followed by desalination (i.e, the expulsion of salt from the ice) processes that might lead to a slow asymptotic decrease of the ice thickness.
Desalination of mushy ice reduces its porosity, which alters the dynamic thermal equilibrium and ice thickness by weakening buoyancy-driven convection within the mushy layer. 
In turn, changes in brine convection and ice thickness affect the further desalination of the ice.
The long-term dynamics of the system can be accurately predicted by a one-dimensional model based on appropriate parameterizations of global heat and mass transfer properties.
Furthermore, within the same theoretical model we explore the ice dynamics across a broader parameter space.
Our findings advance the understanding of the coupled phase-change physics of saline solutions in the presence of convective fluid flows and provide a basis for explaining and predicting real-world phenomena such as the aging of sea ice.

\end{abstract}

\maketitle

\section{I. Introduction}

Freezing and melting problems have received substantial attention in recent fluid dynamics studies \cite{du2024physics,guardone2025aircraft,roach2025physics,huerre2025freezing,wang2025physics}, due to their wide-ranging applications in geoscience and industry.
Specific scenarios include ocean-driven melting of ice shelves \cite{rosevear2021role} and icebergs \cite{cenedese2023icebergs}, the sculpting of scallop patterns on underwater ice surfaces \cite{ristroph2018sculpting, washam2023direct} and the development of energy storage systems with phase-change materials \cite{hu2018close, hu2022rapid, yang2024enhanced, proia2024melting, proia2025heat}.
A frequently posed question in these studies is how the coupling between the solid-liquid phase change and fluid flow depends on the properties of the liquid and solid, as well as the environmental conditions.

Generally, fluid flow can influence heat and mass transport, thereby affecting phase change both locally and globally.
In turn, phase change can also modify the flow by altering the shape of the liquid region and the distribution of temperature and salinity within the fluid.
However, the specific manifestation of this coupled dynamics varies in different situations.
For example, the density anomaly of freshwater introduces non-linear behaviors in the freezing and melting of ice \cite{wang2021growth, wang2021ice, yang2022abrupt, weady2022anomalous}.
In saline water, the coexistence of temperature and salinity gradients may cause double-diffusive convection \cite{yang2023ice, guo2025effects, xu2025buoyancy}.
The presence of a mean flow can also affect the stability of the ice–water interface, leading to rich interfacial morphologies \cite{toppaladoddi2019combined, toppaladoddi2021nonlinear, perissutti2024morphodynamics, perissutti2025time}.
For a finite-size ice object, the local melting rate is related to the orientation of the ice front.
Consequently, the overall melting rate is influenced by its shape, both in background thermal convection \cite{yang2024circular, xu2025aspect} and in turbulent mean flows \cite{yang2024shape, hester2021aspect}.
Other studies have examined the effects of complex conditions, such as inclination \cite{wang2021ice, wang2021equilibrium,  yang2024enhanced},  vibration \cite{fang2024vibration}, system rotation \cite{ravichandran2021melting, ravichandran2022combined}, shear-thinning liquid \cite{hu2023close}, free liquid surface \cite{bellincioni2025melting} and hydrophobic textured surface \cite{hu2025close}.
Moreover, different couplings between a freezing or melting boundary and thermal convection have been shown to induce asymmetric behaviors and multiple equilibrium states of the system \cite{yang2025asymmetric, wang2021equilibrium}.

When it comes to freezing and melting in salty water, an essential characteristic of the formed ice is its porous structure, known as mushy ice \cite{feltham2006sea}.
This porous structure influences both the thermal and mechanical properties of ice \cite{timco2010review}, which are relevant not only for phase change dynamics \cite{worster2015sea} but also for its responses to winds, waves, ocean currents, and other environmental forces \cite{feltham2008sea, roach2025physics}.
Under sufficiently strong buoyant driving, brine convection can occur within the pore structure of mushy ice \cite{anderson2020convective}.
This convection significantly affects the growth of mushy ice layers, leading to the formation of brine channels and the desalination of the ice (i.e, the expulsion of salt from the ice) \cite{wettlaufer1997natural, chung2002steady}.
In our recent studies, we found that brine convection greatly enhances heat transfer in mushy ice and thereby maintains a large ice thickness when the system is in thermal equilibrium \cite{du2023sea}.
We also observed that newly formed mushy ice slowly ages through desalination after the system reaches thermal equilibrium.
This process is characterized by a gradual reduction in porosity, and the ice eventually becomes non-porous \cite{du2025sea}.
While these two studies examined heat and mass transfer separately, the question remains of how these two processes interact and together determine the long-term dynamics.

In this work, we study by means of laboratory experiments the long-term (i.e. asymptotic in time) evolution of the ice layer formed by cooling a salty water liquid layer from the top while heating it from the bottom.
We observe that after a relatively rapid freezing process that leads to the formation of a suspended ice layer in contact with the cold wall, a slower ice thinning (i.e., decreasing of the mean ice thickness) process takes place. 
This is accompanied by a gradual reduction in ice porosity and desalination. 
This affects the thermal equilibrium of the system primarily by weakening convection within the mushy ice, which as a consequence decreases the ice thickness.
By comprehensively modeling heat and mass transfer as well as salt conservation in the system, we can accurately predict the long-term dynamics of the ice layer after it reaches its maximum thickness, using the ice porosity at that specific point as input.
This modeling framework is further extended to explore ice dynamics across a wider range of parameter regimes.

\section{II. Experiments on freezing of saline water solution subjected to convecting fluid motion} 

\subsection{A. Experimental setup}

Icing and melting in a fluid environment are often accompanied by the persistence of thermal gradients that can produce convective motions. We focus here on the case of natural convection as opposed to forced convection. 
This situation is consistent with the thermal settings of the Rayleigh-B\'enard (RB) convection system \cite{esfahani2018basal, wang2021equilibrium, yang2023morphology}, which we will also adopt in the present study.
In the RB system, a temperature difference aligned with the direction of gravity is imposed by a top cold boundary and a bottom hot boundary.
Buoyancy acts on the fluid in the gap of these two parallel boundaries to generate convection.
The RB system has been widely adopted as an ideal model in the investigation of many crucial thermal fluid dynamical problems (see, for example \cite{ahlers2009heat, lohse2010small, lohse2023ultimate}).

\begin{figure*}[!tb]
\centering
\includegraphics[width=1\linewidth]{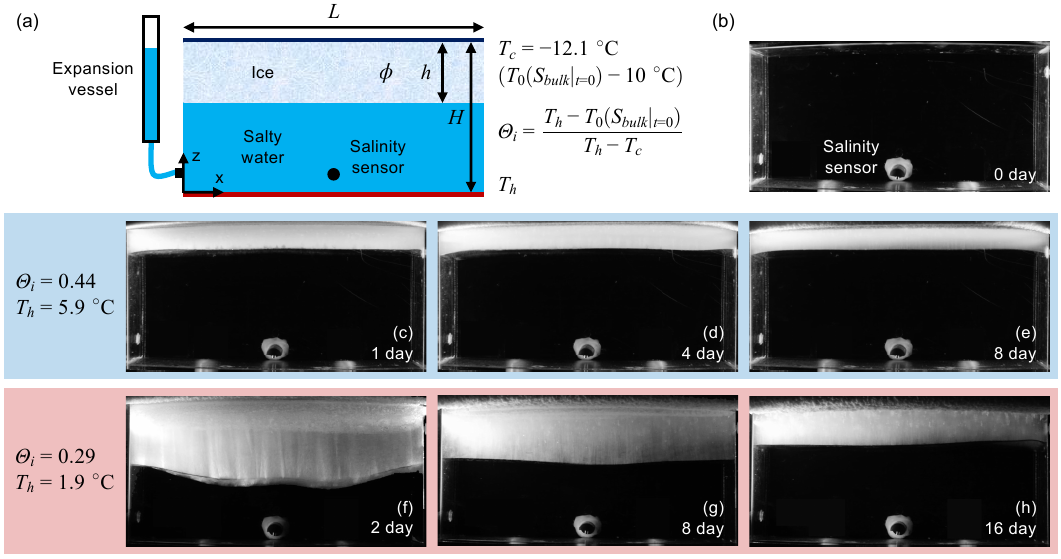}
\caption{Evolution of the mushy ice in salty water. (a) Sketch of the experimental system.
The experiments adopt a cuboidal cell with aspect ratio $L/H=2$.
The cell is initially filled with salty water of salinity $S_{bulk}|_{t=0}\approx3.5\%$.
The cold top plate is maintained at $T_c=-12.1\ ^\circ$C, 10 $^\circ$C below the initial freezing point $T_0(S_{bulk}|_{t=0})$.
The hot bottom plate temperature $T_h$ decides a dimensionless parameter $\Theta_i=(T_h-T_0(S_{bulk}|_{t=0}))/(T_h-T_c)$.
A salinity sensor is embedded near the bottom plate.
An expansion vessel is connected to compensate for the volume change.
(b) Image at the start of the experiment.
(c-e) Images of the ice layer for $\Theta_i=0.44$ ($T_h=5.9$ $^\circ$C) at day 1 (c), day 4 (d) and day 8 (e).
(f-h) Images of the ice layer for $\Theta_i=0.29$ ($T_h=1.9$ $^\circ$C) at day 2 (f), day 8 (g) and day 16 (h).
}
\label{fig1}
\end{figure*}

\begin{figure*}[!tb]
\centering
\includegraphics[width=1\linewidth]{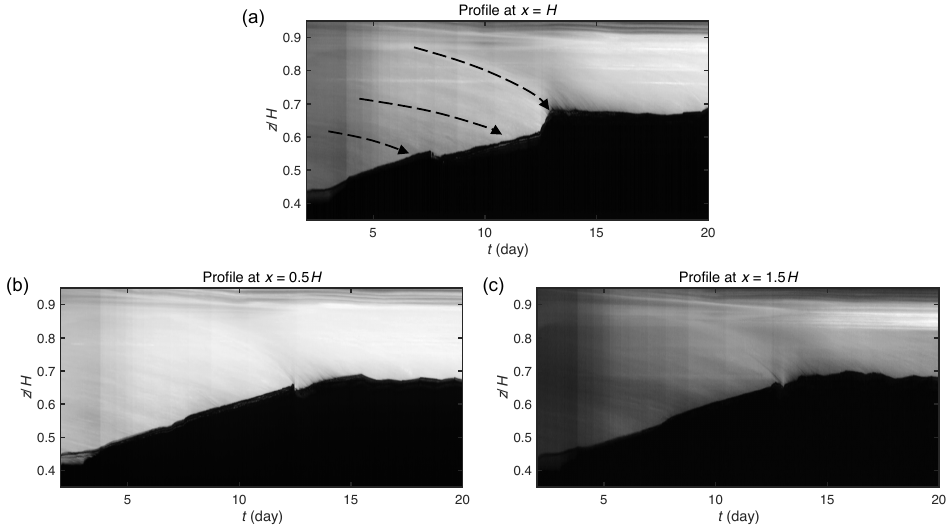}
\caption{Space-time diagram for the ice layer profile at different horizontal positions $x=H$ (a), $x=0.5H$ (b) and $x=1.5H$ (c), in the experiment with $\Theta_i=0.29$. 
The dashed arrows in (a) are drawn to guide the eye for the stripes.
}
\label{fig2}
\end{figure*}

The experiments are performed in the RB system shown in Fig.~\ref{fig1}a.
The system is composed of a cuboidal cell with length $L=0.24$ m in the $x$ direction, height $H=0.12$ m in the $z$ direction and width $W=0.06$ m (see Figs.~\ref{fig1}a,b).
In this geometry, the flow is quasi-two-dimensional and in the $x-z$ plane.
The cell is initially filled with salty water (aqueous sodium chloride solution) of salinity $S_{bulk}|_{t=0}\approx3.5\%$, corresponding to an initial freezing point of $T_0(S_{bulk}|_{t=0})=-2.1\ ^\circ$C.
This salinity is approximately the average salinity of the oceans.
During the experiments, the cold top plate of the cell is maintained at $T_c=-12.1\ ^\circ$C.
The effect of temperature is examined by varying the temperature of the hot bottom plate $T_h$, which is also maintained constant in time during each experiment.
To quantify different levels of bottom thermal forcing, we define a dimensionless parameter as
\begin{equation}
    \Theta_i=\frac{T_h-T_0(S_{bulk}|_{t=0})}{T_h-T_c}.
\end{equation}
This parameter represents the temperature difference in the initial liquid divided by the total temperature difference in the system.
$\Theta_i=0$ means that the highest temperature in the system, which is the temperature of the hot plate, is equal to the initial freezing point.
The entire system would freeze if the changes in salinity during the experiments were negligible.
On the other hand, $\Theta_i\xrightarrow{}1$ means that the difference between the cold plate temperature and the freezing point is negligible compared to the temperature difference in the liquid.
In this case, no ice layer can form.
In the present work, we focus on two different cases: (1) $\Theta_i = 0.44$ ($T_h = 5.9\ ^\circ$C, see Movie S1 in the supplemental materials) and (2) $\Theta_i = 0.29$ ($T_h = 1.9\ ^\circ$C, see Movie S2 in the supplemental materials).
The ambient temperature is controlled around $0.5(T_c + T_h)$ to minimize heat transfer through the side walls.

In the experiments, the mushy ice layer forms beneath the cold top plate.
The evolution of this ice layer is monitored with a camera, from which its mean thickness $h$ can be determined (see Figs.~\ref{fig1}b–h).
In addition, a salinity sensor is embedded near the bottom plate.
Assuming that the salt is well mixed, resulting in a uniform salinity throughout the liquid, the sensor reading corresponds to the average value.
Using this value, the mean salinity and temperature of the brine in the mushy ice can be estimated by incorporating appropriate assumptions on the local temperature, salinity and porosity profiles in ice (see more in Appendix A and B).
Then, the mean porosity of the ice layer, $\phi$, can be calculated by applying salt conservation, as discussed in Appendix B below.
Finally, since the density of ice is roughly one-tenth lower than that of salty water, an expansion vessel is connected to the main cell to compensate for the volume change during the experiments.

\subsection{B. Qualitative observations}

Figures \ref{fig1}c–e and f–h show images of the ice layer taken during the experiments when $\Theta_i = 0.44$ and $0.29$, respectively.
The ice layer is opaque because salty water freezes into mushy ice with an internal porous structure.
In both cases, the ice layer grows rapidly over 1 to 2 days to reach maximum thickness (see Figs.~\ref{fig1}c,f).
However, the ice layer behaves very differently afterward.
When the bottom plate temperature is high ($\Theta_i = 0.44$, Figs.~\ref{fig1}c–e), the ice thickness remains nearly constant for the remainder of the experiment.
In contrast, when the bottom plate temperature is low ($\Theta_i = 0.29$, Figs.~\ref{fig1}f–h), the ice layer thins (i.e., the mean ice thickness $h$ decreases) significantly over the following two weeks. 
This behavior was not noticed in our previous study with the same setup \cite{du2023sea}, because the observation time was limited to the first initial growth stage. The long observation time constitutes the focus of the present work.
The color of ice layer in these images does not have a uniform shade of white, but has light and dark patches.
These patterns are likely to be related to the non-uniform porous structure in the ice interior.
They appear to move slowly downward to the ice front, see the supplementary videos.
This indicates the migration of the pores structures.
This is analogous, although rotated 90 degrees, to what we observed for the case of freezing a saline solution in a vertical convecting system \cite{du2025sea}.

To better visualize the evolution of the mushy ice layer in saline water, the snapshots shown in Fig.~\ref{fig1}(f–h) are re-plotted as space-time diagrams at three horizontal positions: $x = H$ (a), $0.5 H$ (b) and $1.5 H$ (c) in Fig.~\ref{fig2}. These diagrams reveal two dominant processes: (i) the overall melting of the ice layer, evidenced by the progressive shrinkage of the white region, and (ii) the migration of the pore structure within the ice, inferred from the inclined stripes highlighted in Fig.~\ref{fig2}(a).
All three space-time diagrams are quite consistent.
Therefore, it can be inferred that the pore migration is similar at different horizontal locations.
In the space-time diagrams, the slope of these strips reflects the velocity.
A common feature of the strips is that they curve slightly downward.
This suggests that pore migration accelerates as the pore approaches the ice front.
In addition, the slopes of the strips are larger after around day 10 than before, indicating faster pore migration.
While estimating the migration velocity of the pore structures with space-time diagrams may not be very accurate, it is basically certain that the migration velocity in the experiment with $\Theta_i=0.44$ is in the order of $10^0$ mm/day.

\begin{figure*}[!tb]
\centering
\includegraphics[width=1\linewidth]{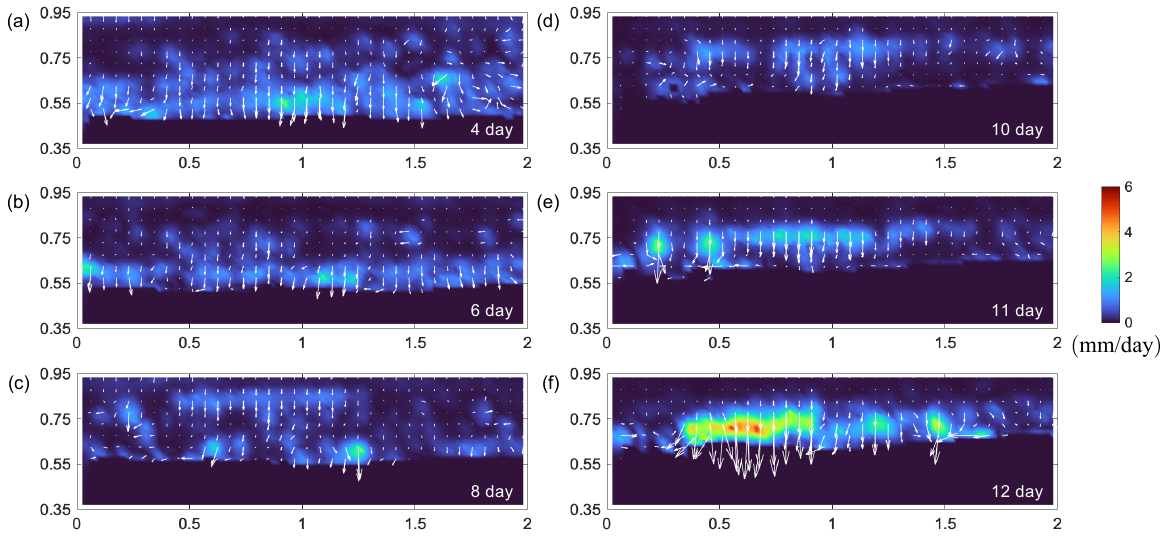}
\caption{Velocity fields in the ice layer at $t=4$ day (a), 6 day (b), 8 day (c), 10 day (d), 11 day (e) and 12 day (f) for $\Theta_i=0.29$.
The color contour shows the magnitude of the migration velocity.
The direction and length of the white arrows indicate the direction and magnitude of the migration velocity.
}
\label{fig3}
\end{figure*}

When the patterns on the ice layer display sufficient features, the migration of the pore structure can be measured by examining the correlation between two consecutive images \cite{adrian1991particle}.
Fig.~\ref{fig3} shows the velocity fields in the ice layer at different times in the experiment with $\Theta_i=0.29$.
The velocity fields are computed with the two images taken 3 hours before and after the labeled time.
A window size of 32$\times$32 pixels is adopted, with a 50$\%$ overlap (for reference, the area shown in Figs.~\ref{fig3} is of 1700$\times$560 pixels).
Despite restrictions such as the limited distinguishability of the pattern features, the continuous deformation of pore structures, and the image quality, the velocity fields still clearly show that the pore structure migration is mainly in the downward direction.
These restrictions also make the computations in most of the windows invalid.
Therefore, we identify the typical velocity as the average of the largest 5$\%$ velocities in each velocity field.
For day 4 to day 10 (Figs.~\ref{fig3}a-d), the typical velocities are $1.37$, $1.14$, $1.03$ and $0.88$ mm/day; for day 11 and day 12, they are $1.52$ and $2.96$ mm/day (Figs.~\ref{fig3}e-f).
It is noted that, while difficult to visualize with image correlation due to the image quality, such migration with a similar velocity is also observable in the experiment with $\Theta_i=0.44$, see the supplementary video.
Image correlation has its limitations; however, some qualitative results can still be obtained.
Pore migration is generally faster and more easily observed in the lower part of the ice layer.
The typical migration velocities slightly decrease before day 10 and then increase markedly on days 11–12.
The typical migration velocities are on the order of millimeters per day.
These findings are consistent with the results of the space-time diagrams.
Specifically, the space-time diagrams (Fig.~\ref{fig2}) show how the pores migrate over time at different horizontal positions, while the velocity fields (Fig.~\ref{fig3}) show how the migration velocity varies in space at different times.
Together, they provide consistent information on the migration of the pore structure.

\begin{figure*}[!tb]
\centering
\includegraphics[width=1\linewidth]{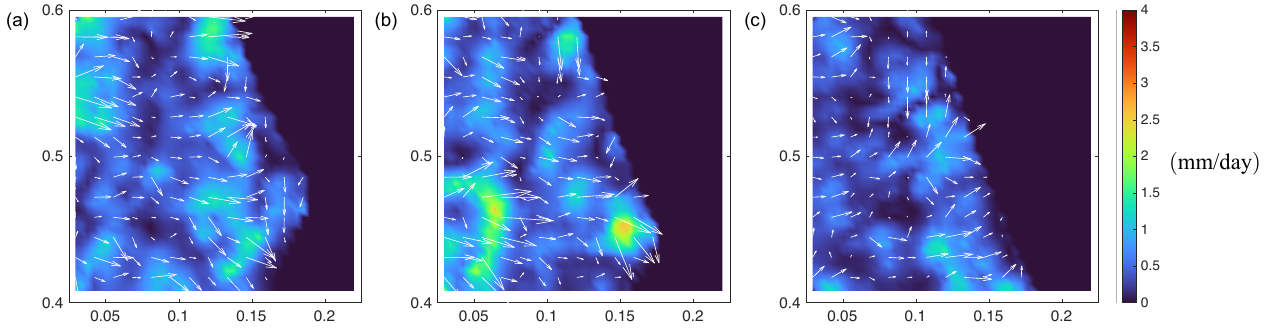}
\caption{Velocity fields in the ice layer at $t=2$ day (a), 3 day (b), 4 day (c) in the experiment of long-term ice evolution in a vertical convection system with $\Theta_i=0.5$.
The original images are from Ref. \cite{du2025sea}.
The color contour shows the magnitude of the migration velocity.
The direction and length of the white arrows indicate the direction and magnitude of the migration velocity.}
\label{fig4}
\end{figure*}

We also applied image correlation to analyze the results of our recent experiment \cite{du2025sea} on long-term ice evolution in a vertical convection system, changing only the window size to 16$\times$16 pixels.
The height of the system was $H' = 0.24$ m and the superheat was $\Theta_i = 0.5$.
Fig.~\ref{fig4} shows the velocity fields in the region $0.025H' \le x \le 0.225H'$ and $0.4H' \le z \le 0.6H'$ at different times.
The velocities are primarily directed from inside the ice toward the bulk salty water, and seem to turn toward the perpendicular direction near the ice front.
The typical velocities are 1.31, 1.6 and 1 mm/day, which are of the same order of magnitude as the result (3.0 mm/day) obtained from the space-time diagram reported in Ref.~\cite{du2025sea}.
This consistency further validates the results of image correlation in the present work.

Returning to the top-down freezing experiments in this work, when a high bottom temperature provides strong thermal forcing to the system ($\Theta_i = 0.44$), the ice layer remains nearly constant in thickness once its maximum is reached.
In contrast, under low bottom temperature conditions ($\Theta_i = 0.29$), the ice layer thins considerably over the long term.
However, within the mushy ice in both cases, similar pore structure migrations are observed.
These results raise several interesting questions.
Why does the ice layer behave so differently under different temperature conditions?
How do changes inside the mushy ice affect the global dynamics of the ice front?

\section{III. Dynamics of the long-term ice evolution}

Answering these questions requires understanding the long-term (i.e. asymptotic) dynamics of ice layer evolution. The ice dynamics is controlled by heat and mass transfer in the system. At the beginning of the experiment, the moving ice front is governed by the difference between the heat fluxes across the ice layer and the bulk salty water, which is described by the Stefan interface condition \cite{gupta2017classical}.
During ice growth, salt cannot be incorporated into the forming ice and instead remains in the unfrozen liquid \cite{vrbka2005brine}.
The time scale for heat transfer can be estimated as $t_\kappa=(h^*)^2/\kappa_{brine}$ where $(h^*)^2$ is the typical length scale of the ice and $\kappa_{brine}$ is the thermal diffusivity in the brine.
The one for phase change can be estimated as $t_{ste}=(h^*)^2Ste/\kappa_{ice}$ where the Stefan number $Ste=\mathcal{L}/(c_{pice}(T_c-T_0))\approx17$ is the ratio between latent heat and sensible heat in ice and $\kappa_{ice}$ is the thermal diffusivity of ice.
The one for mass transfer can be estimated as $t_D=(h^*)^2/D_{brine}$, where $D_{brine}$ is the solutal diffusivity in the brine.
As the Lewis number $Le_m=t_D/t_\kappa=\kappa_{brine}/D_{brine}\approx200$ and $t_{ste}/t_\kappa=(\kappa_{brine}/\kappa_{ice})Ste\approx1.6$, it can be inferred that the time scales for heat transfer and phase change are much shorter than those for mass transfer. 
As a result, excess salt becomes trapped inside the mushy ice as highly concentrated brine.
As the ice layer thickens, the heat flux difference across the ice front decreases.
The system first reaches thermal equilibrium when the maximum ice thickness is attained \cite{du2023sea}.

Over the next long period of time after the initial rapid ice growth, the ice layer continues to evolve.
In this work, we focus particularly on the ice dynamics in this period.
After the initial rapid ice growth, the ice front moves very slowly or may even remain stationary.
Thermal equilibrium can be inferred to persist during the long-term evolution of the ice.
Meanwhile, the temperature and salinity of the brine also satisfy the local thermal equilibrium (see \cite{notz2009desalination}).
In other words, the freezing/melting point at the local brine salinity equals the local temperature.
With the local temperature increasing from the cold top plate to the ice front, the local salinity decreases in the $-z$ direction.
This induces a downward transfer of salt, which is a relatively slow process until the asymptotic solutal equilibrium is reached.

\subsection{A. Migration of pore structures}

On the scale of the pores within the mushy ice, salt is transported from the colder side to the hotter side.
As mentioned above, the thermal time scale is much shorter than the solutal one.
Hence, freezing occurs on the colder side of the pores, and melting occurs on the hotter side, maintaining local thermal equilibrium.
This results in the observed migration of the pore structures.
It should be noted that pore structures and their migration are more easily observed for isolated pores, which have a clear contrast with the surrounding ice, rather than for vertically extended brine channels.
Brine convection is unlikely to have a strong influence on this migration.
The pore migration can be effectively modelled as a diffusion process, and the migration velocity for isolated pores can be estimated using the apparent velocity for vertical salt diffusion \cite{notz2009desalination}.
Assuming that the salinity near the ice front is approximately the salinity measured by the sensor, the apparent velocity can be estimated, as shown in Appendix A.
It is approximately 3.0 mm/day in the experiment with $\Theta_i = 0.44$ and 0.7–1.6 mm/day in the experiment with $\Theta_i = 0.29$, both on the order of $10^0$ mm/day.
While these theoretical and experimental estimates are approximate, their consistency provides confidence in the above interpretation of the heat and mass transfer processes in the system.
The estimated $v_D$ values are generally lower than the observed migration velocities.
This suggests the existence of additional factors that could accelerate salt diffusion, such as the nonuniform distribution of ice porosity.

\subsection{B. Evolution of the global properties of the system}

Furthermore, noting that the ice layer varies little in the horizontal ($x$) direction and the width (perpendicular to the page) direction, we aim to understand the long-term ice dynamics and other system properties quantitatively by modeling heat and mass transfer in the vertical ($z$) direction.
As discussed above, the system behavior is controlled by quasi-equilibrium heat transfer, the desalination of mushy ice, and the conservation of salt in the system.
By quantifying these three processes, the mean ice thickness, mean ice porosity, and mean bulk salinity can be predicted as functions of time, while properly parameterizing other properties needed in the model.
Compared to Refs.~\cite{du2023sea, du2025sea}, which examine heat or mass transfer independently and rely on inputs of the time series of experimental measurements, comprehensively modeling these processes reduces the input to a single value of the ice porosity at maximum ice thickness as a definite condition and several empirical parameters on the ice structure and transport properties.
Here, we do not repeat the detailed equations, but refer to Appendix B for more information on the model.

\begin{figure*}[!tb]
\centering
\includegraphics[width=1\linewidth]{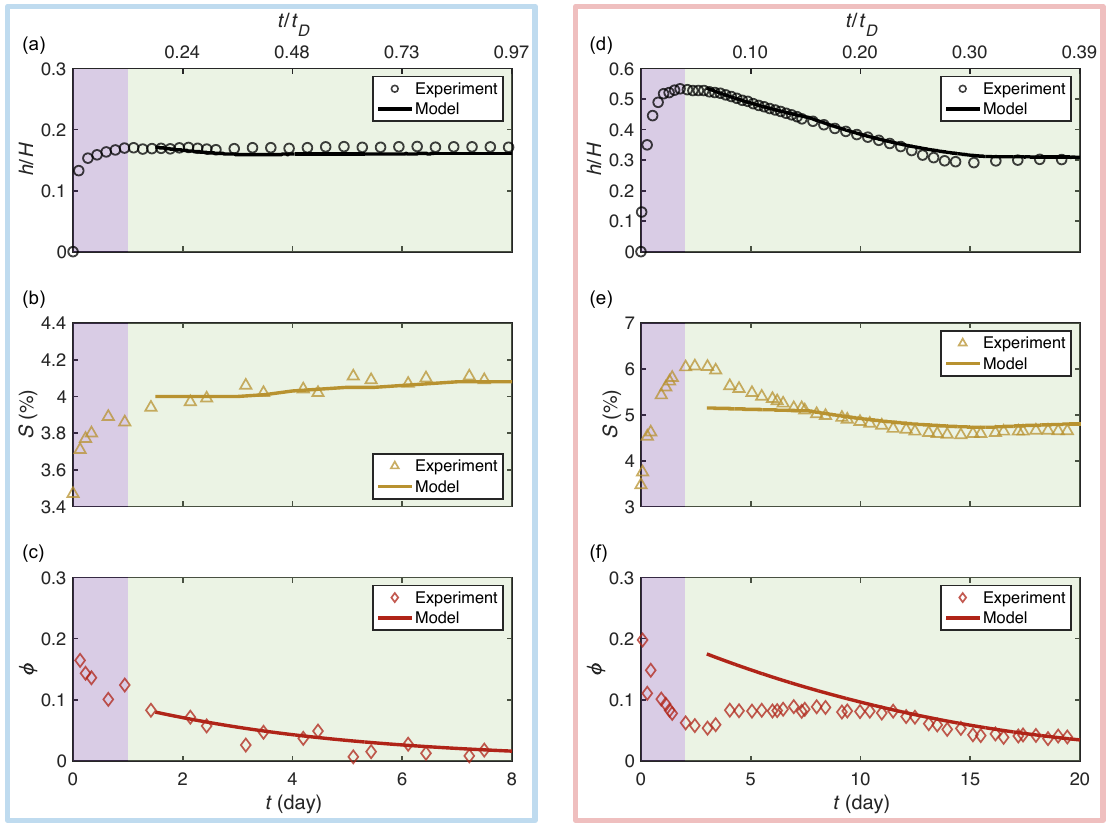}
\caption{Experiment results (symbols) and theoretical predictions (lines) on mean ice thickness $h/H$ (a,d, black), salinity measured by the sensor $S$ (b,e, yellow) and mean ice porosity $\phi$ (c,f, red) as function of time.
The ice porosity in the experiments (red diamonds in c,f) is calculated with mass conservation inputting the measured ice thickness and salinity.
The top axes exhibit the dimensionless time $t/t_D$, where $t_D=(h^*)^2/D_{brine}$ is the solutal diffusive time scale, $h^*$ is the spatial-temporal mean ice thickness and $D_{brine}$ is the salt diffusivity.
(a-c) Experiment results and theoretical predictions for $\Theta_i=0.44$, with $t_D\approx8.3$ days. 
(d-f) Experiment results and theoretical predictions for $\Theta_i=0.29$, with $t_D\approx50.7$ days. 
}
\label{fig5}
\end{figure*}

Fig.~\ref{fig5} shows the experimental results and theoretical predictions for the three major global properties of the system discussed above.
In the experiment with $\Theta_i = 0.44$, the ice layer remains at its maximum thickness after initial growth (black dots in Fig.~\ref{fig5}a).
The salinity measured by the sensor increases in the long term (yellow triangles in Fig.~\ref{fig5}b).
Calculated using the measured ice thickness and salinity, with mass conservation of salt (Eq.~\ref{eq:mass}) incorporating appropriate assumptions on the mean salinity and temperature of the brine in ice (see the detailed assumptions in Appendix B), the mean ice porosity is found to generally decrease with time (red diamonds in Fig.~\ref{fig5}c).
The proposed model predicts the temporal evolutions of these three properties well (lines in Figs.~\ref{fig5}a-c).
This strongly supports the idea that the mechanism described above captures the long-term ice dynamics: The system is in thermal equilibrium, and ice desalination leads to an increase in the bulk salinity and further freezing within the mushy ice.

In the experiment with $\Theta_i = 0.29$, the ice layer thins considerably after the initial growth before reaching its asymptotic final thickness (black dots in Fig.~\ref{fig5}d).
The salinity measured by the sensor decreases (yellow triangles in Fig.~\ref{fig5}e).
However, the calculated mean ice porosity decreases to a very low value during initial ice growth, slightly increases and remains nearly constant over the following few days, and then decreases again in the latter half of the experiment (red diamonds in Fig.~\ref{fig5}f).
This unusual trend in ice porosity is probably due to an overestimation of the bulk salinity.
Ref.~\cite{wettlaufer1997natural} suggests that the salt expelled from the mushy ice may be carried to the bottom of the system by plumes when brine convection occurs in the ice.
This can lead to the establishment of double-diffusive convection and the formation of a highly saline region near the bottom.
As a result, the salinity measured by the sensor is likely higher than the mean bulk salinity.
Using the measured value to estimate the mean bulk salinity in the mass conservation calculation causes an underestimation of the mean ice porosity.
It is noted that the non-uniform salinity in the bulk salty water also affects the estimations of the mean values or gradients of salinity or temperature in the theoretical modelling.
However, during the initial growth of ice, the salinity at the ice front must be larger than about $3.5\%$, the mean bulk salinity is about $5\%$ (Fig.~\ref{fig5}e), while the cold plate temperature is $-12.1$ $^\circ$C corresponding to a local salinity of about $16.1$ $^\circ$C.
It can be estimated that the maximum relative error in these estimation is no more than $13\%$ and acceptable (see Fig.~\ref{fig8} in Appendix C for the freezing/melting point of NaCl solution as function of salinity \cite{hall1988freezing}).
During the following long-term evolution, the decrease of ice thickness is very slow, hence the convection in the bulk salty water should be able to mix the melt water.
The salinity at the ice front is not expected to decrease much further.
As the salt mixes in the bulk, the difference between the measured salinity and the mean bulk salinity decreases, making the calculated porosity in the latter half of the experiment more reliable.
The model, based on the previous theoretical analysis, successfully predicts the temporal evolution of the bulk salinity and mean ice porosity in the latter half of the experiment, and accurately predicts the mean ice thickness throughout the experiment from $t = 3$ days (lines in Figs.~\ref{fig5}d-f).
This agreement, particularly in ice thickness, supports the above discussion and the proposed mechanism of ice dynamics.
Similar to the case with $\Theta_i = 0.44$, mushy ice desalination leads to a decrease in ice porosity when $\Theta_i = 0.29$.
The combined effects of mushy ice desalination and ice melting result in only slight changes in the bulk salinity.
The slight changes in the mean bulk salinity and the corresponding freezing/melting point very limited compared to the mean salinity in ice as well as the salinity and temperature differences across the ice layer and the bulk salty water layer.
Hence, they are are not expected to have a considerable influence on the ice evolution.

\begin{figure*}[!tb]
\centering
\includegraphics[width=1\linewidth]{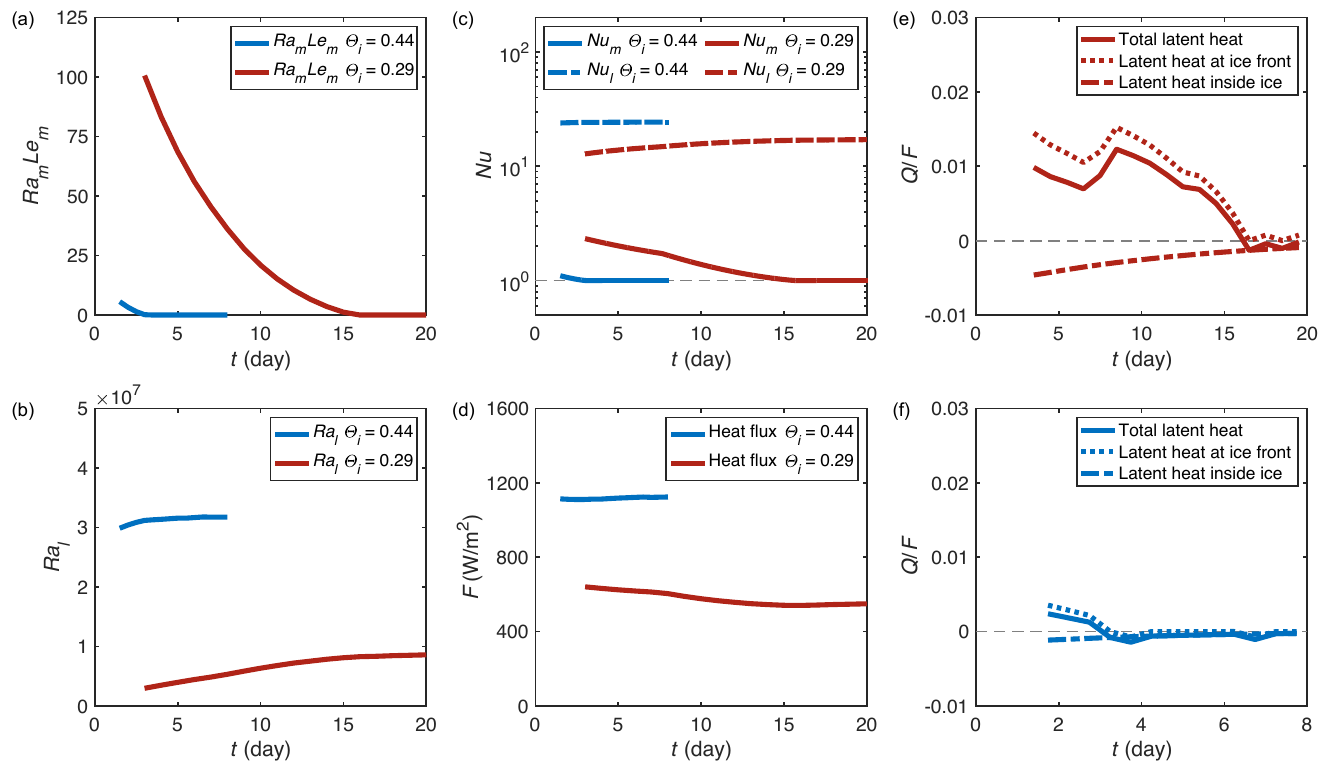}
\caption{Theoretical results on the buoyancy intensity and heat transfer properties as function of time for $\Theta_i=0.29$ (red lines) and $\Theta_i=0.44$ (blue lines).
(a) Dimensionless buoyancy intensity in mushy ice $Ra_mLe_m$.
The Rayleigh number $Ra_m$ is multiplied by the Lewis number $Le_m$ because the buoyancy in mushy ice is primarily induced by salinity gradient in the present study.
(b) Dimensionless buoyancy intensity in bulk salty water $Ra_l$.
(c) Nusselt numbers (dimensionless heat transfer efficiency) in mushy ice ($Nu_m$, solid lines) and bulk liquid ($Nu_l$, dash-dotted lines).
The gray dashed line sketches $Nu=1$ (diffusive heat transfer).
(d) Heat fluxes through the system $F$.
(e,f) Ratio between the latent heat absorption $Q$ and heat flux $F$ in the system.
The solid lines show the total latent heat absorption, the dotted lines show the latent heat absorption at the ice front, and the dash-dotted line show the latent heat absorption inside the mushy ice.
The gray dashed line sketches $Q=0$, with $Q>0$ indicating overall melting and $Q<0$ indicating overall freezing.
}
\label{fig6}
\end{figure*}

Furthermore, theoretical modeling shows that the desalination of mushy ice affects the flow dynamics and heat transfer in the system (Fig.~\ref{fig6}), which determines the distinct ice front dynamics observed in the two experiments.
The flow dynamics can be evaluated with the dimensionless buoyancy intensities in mushy ice and in bulk salty water.
According to equation of state for salty water \cite{gebhart1977new}, the buoyancy in mushy ice is primarily induced by salinity gradient.
Hence, the effective controlling parameter for buoyancy-driven brine convection in ice is the Rayleigh number $Ra_m$ multiplied by the Lewis number $Le_m$.
The effective controlling parameter for convection in bulk salty water is the Rayleigh number $Ra_l$.
Their exact definitions are provided in Appendix B.
As the effective buoyancy intensities vary with time, the temporal evolutions of $Ra_mLe_m$ and $Ra_l$ are reported here in Fig.~\ref{fig6}(a,b) instead of introduced when describing the experimental setup.
The heat transfer in the system can be evaluated with the dimensionless heat transfer efficiencies, the Nusselt numbers in mushy ice $Nu_m$ and in bulk salty water $Nu_l$ (Fig.~\ref{fig6}c), as well as the dimensional heat flux $F$ and latent heat absorption and release $Q$ (Fig.~\ref{fig6}d-f).

When $\Theta_i = 0.44$, the strong thermal forcing from the bottom plate suppresses the ice thickness.
With low ice thickness and low ice porosity, the buoyancy in mushy ice is weak (blue line in Fig.~\ref{fig6}a).
As a result, convection and its enhancement of heat transfer in ice are nearly negligible ($Nu_m \approx 1$), in contrast to the strong convection in the bulk salty water (blue lines in Fig.~\ref{fig6}b,c).
The desalination of the ice leads primarily to a decrease in ice porosity, which in turn affects the effective thermal diffusivity in the mushy ice.
However, this influence is expected to be marginal, as the ice thickness and heat flux ($F$, see Eq.~\ref{eq:heat}) through the system remain stable throughout the experiment, as shown by the blue lines in Fig.~\ref{fig5}(a) and Fig.~\ref{fig6}(d).

The relationship between heat and mass transfer becomes more complex when $\Theta_i = 0.29$.
In this case, the low bottom temperature allows the ice to grow thicker.
As the thickness and porosity of the ice layer increase, the buoyancy becomes high enough to drive the brine convection, which enhances heat transfer in the mushy ice and further facilitates ice growth.
As a result, the ice layer initially grows to a considerable thickness.
In the long term, convection in the mushy ice also accelerates the ice desalination and the decrease in ice porosity.
This is evident when comparing the duration of the experiment with the solutal diffusion time scale, $t_D=(h^*)^2/D_{brine}$, where $h^*$ is the spatial-temporal mean ice thickness and $D_{brine}$ is the salt diffusivity, as shown by the top axes in Fig.~\ref{fig5}.
The reduction in ice porosity weakens the buoyancy in mushy ice, as shown by Eq.~\ref{eq:Ram} (red line in Fig.~\ref{fig6}a).
Consequently, convection weakens and heat transfer efficiency in the mushy ice decreases (red solid line in Fig.~\ref{fig6}c), and the ice layer gradually thins over time (Fig.~\ref{fig5}d).
Eventually, convection in the mushy ice ceases and the ice layer reaches its asymptotic final thickness.
Assisted by thinning of the ice, buoyancy in the bulk salty water increases (red line in Fig.~\ref{fig6}b), strengthening the bulk convection before the system reaches its asymptotic final state (red dash-dotted line in Fig.~\ref{fig6}c).
As a result, the heat flux through the system decreases (Fig.~\ref{fig6}d).

Melting at the ice front absorbs latent heat $Q_1=-\rho_{ice}\mathcal{L}(1-\phi)(dh/dt)$, as indicated by the dotted lines in Figs.~\ref{fig6}(e,f). 
In contrast, freezing in the pores of the mushy ice releases latent heat $Q_2=\rho_{ice}h\mathcal{L}(d\phi/dt)$, as indicated by the dashed-dotted lines in Figs.~\ref{fig6}(e,f).
Here, $Q$ denotes the latent heat absorption per unit cross-sectional area, $\rho_{ice}$ is the ice density and $\mathcal{L}$ is the latent heat of ice-water phase change. 
The sum of these two processes ($Q_1+Q_2$) results in total latent heat absorption and release, as shown by the solid lines in Figs.~\ref{fig6}(e,f).
As the system approaches its asymptotic final state, the phase change slows, and the latent heat generally decreases.
Throughout the experiments, after the initial ice growth, the latent heat absorption and release are much smaller than the heat flux through the system.
This further supports the idea that the system remains in thermal equilibrium during this process.

To briefly summarize, after the initial rapid growth of ice, the system remains in quasi-thermal equilibrium over the long term.
Desalination of the mushy ice leads to a decrease in ice porosity, which in turn, affects heat transfer and the ice layer thickness.

\section{IV. Ice evolution in wider parameter regimes}

\begin{figure*}[!htb]
\centering
\includegraphics[width=1\linewidth]{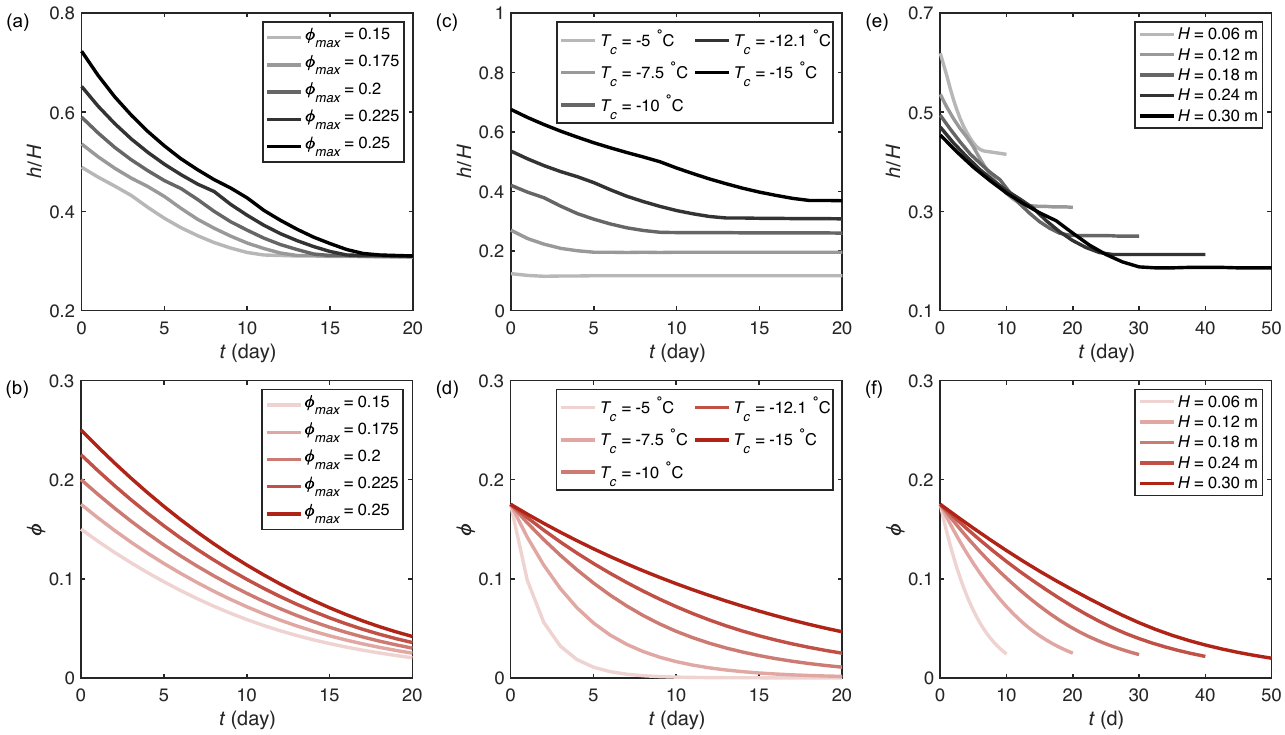}
\caption{Theoretical predictions on the temporal evolution of mean ice thickness $h/H$ (black lines) and mean ice porosity $\phi$ (red lines) with different maximum ice porosities $\phi_{max}$ (a,b), different cold plate temperatures $T_c$ (c,d) and different system heights $H$ (e,f).
The other parameters and coefficients are the same as those in Fig.~\ref{fig5}(d-f) (the case with $\Theta_i=0.29$) except for the investigated one.
}
\label{fig7}
\end{figure*}

The results above show that the 1D quasi-equilibrium model is able to predict the ice dynamics by inputting the ice porosity at the maximum ice thickness, properly parameterizing the global heat and mass transfer with several empirical parameters on the ice structure and transport properties.
It is therefore tempting to explore how ice evolves over a wider range of system settings and initial conditions compared to the ones of our experimental setup.
In this section, all parameters are kept the same as those used in the case with $\Theta_i = 0.29$, except for the parameter being investigated.
Of course, the results in this section are based on simplifications and assumptions, which may differ from the real scenarios.
However, they still offer a general trend of how the changes in the environmental conditions may affect the system behaviors, which is of practical importance for geophysical and industrial applications.

First, the ice porosity at maximum ice thickness $(\phi_{max})$ is used as a free condition in the present model.
It is related to the coupled, transient heat and mass transfer and phase change during ice growth (see, for example, Refs. \cite{wettlaufer1997natural, peppin2008steady}).
Figs.~\ref{fig7}(a,b) examine the effect of initial ice porosity on subsequent ice evolution.
Higher initial porosity leads to a higher initial ice thickness, as it allows for stronger convection in the mushy ice.
The duration before the ice layer reaches its asymptotic final thickness is longer.
Notably, the lines in Fig.~\ref{fig7}(a,b) overlap each other by translation in time, indicating that the evolution of ice with different initial porosities does not exhibit a memory effect of the initial state.
This is because the ice thickness and bulk salinity at any time depend on the ice porosity at that moment, and the rate of change of the porosity depends on the ice thickness and bulk salinity at that time.

The cold plate temperature $T_c$ makes a difference as well.
In our experiments, the initial salinity is chosen as the average ocean salinity (3.5\%) and the cold plate temperature is chosen as $-12.1$ $^\circ$C, so that the initial temperature difference in the ice $T_c-T_0(S_{bulk}|_{t=0})$ is about $10$ $^\circ$C (see Fig.~\ref{fig8} in Appendix C \cite{hall1988freezing}), which is a typical value in winter.
Obviously, the temperature difference in ice increases with the decrease of $T_c$.
The ice thickness and the brine convection intensity both increase, and the latter is evident from the larger difference between the maximum ice thickness and the asymptotic final ice thickness (Fig.~\ref{fig7}c).
The influence on the ice desalination is more complicated, as the local salinity at the cold plate is determined by $T_c$, as shown by Fig.~\ref{fig8}.
Hence, the mean salinity in mushy ice, the salinity difference across the ice layer, the brine convection intensity and the ice thickness all change with $T_c$.
Despite this, Fig.~\ref{fig7}(d) shows that within the investigated regime, the decrease of ice porosity slower with increasing $T_c$.
This agrees with the previous results that the typical time scale of the long-term ice evolution is the mass diffusion time scale $t_D=(h^*)^2/D_{brine}$, and the typical length scale $h^*$ decreases increasing $T_c$.

Another key parameter that deserves consideration is the height of the system $H$.
The length scale in realistic sea ice settings is usually much larger than in our laboratory experiments.
Fig.~\ref{fig7}(e,f) shows how the dimensionless ice thickness and ice porosity vary with time when the system has different heights.
In Fig.~\ref{fig7}(e), the maximum dimensionless ice thickness appears to decrease with increasing system height.
This observed trend is determined by the scaling of $Nu_m$ and $Nu_l$ with typical length scales.
In the investigated parameter regime, diffusion still contributes significantly to heat transfer in mushy ice and affects the scaling of $Nu_m$.
When the system height is larger, the trend in the maximum dimensionless ice thickness can change.
Additionally, the asymptotic final dimensionless ice thickness decreases with increasing system height.
This is because the ice layer becomes nearly nonporous in the asymptotic final state and $Nu_m = 1$, while $Nu_l$ increases with the typical thickness of the bulk liquid layer.
Meanwhile, it takes more time for the system to reach its asymptotic final state as the system height increases.
This is the combined result of the solutal diffusive time scale and the acceleration of mass transfer by convection in mushy ice.
The solutal diffusive time, which scales with the square of the typical ice thickness, increases more than the mass transfer efficiency in mushy ice as the typical length scale increases.

\section{V. Conclusion and outlook}

By combining experiments and theoretical analyses, we investigate the long-term dynamics of salty water freezing in a Rayleigh-B\'enard convection system.
After the ice reaches its maximum thickness, the dynamics of the ice front can vary depending on the degree of coupling between mass transfer and quasi-thermal equilibrium in the system.
Desalination of the mushy ice, characterized by the migration of the pore structures, leads to a slow decrease in ice porosity, and this porosity decline affects the ice front position mainly by influencing the brine convection in ice.
When the bottom thermal forcing is strong, the maximum ice thickness is small and the heat transfer in the ice is essentially diffusive.
The reduction in ice porosity has minimal impact on heat transfer, and the ice layer remains at its maximum thickness in the long term.
In contrast, weak bottom thermal forcing facilitates both ice growth and buoyancy convection in the mushy ice.
The reduction in ice porosity weakens the convection in mushy ice, leading to shifts in the thermal balance of the system, and the ice gradually thins to an asymptotic final thickness.
In turn, the changes in the strength of brine convection and in the ice thickness affects the ice desalination process.

By comprehensively examining the quasi-equilibrium heat transfer, the desalination of mushy ice, and the mass conservation of salt in the system, we accurately predict the evolution of ice thickness, ice porosity, and mean bulk salinity, using appropriate parameterizations of the global heat and mass transfer properties.
Furthermore, we explore the system dynamics when the ice porosity at maximum ice thickness or the height of the system varies.
It is found that the ice porosity does not exhibit memory effects on the ice evolution, while the height of the system affects both the asymptotic final dimensionless ice thickness and the time required to reach the asymptotic final state.

The findings of this work enhance our understanding of the coupled physics of fluid flow, heat and mass transfer, and phase change during the long-term evolution of a porous medium.
They provide a theoretical reference for explaining and predicting a variety of geophysical and industrial processes, such as the aging of sea ice.
Clearly, real-world scenarios can be much more complicated.
For example, in the long-term evolution of sea ice, the detailed structure of mushy ice may influence brine convection and transfer properties inside mushy ice.
Various flow conditions can occur beneath sea ice \cite{hewitt2020subglacial}, such as shearing flow \cite{perissutti2024morphodynamics, perissutti2025time}, which affect the parameterizations of heat and mass transfer \cite{blass2020flow, scagliarini2014heat}.
There may be complex heat and mass exchanges with the environment, such as unstable environmental conditions \cite{yang2024melting}, solar radiation \cite{yang2023bistability} and mixing with the ambient ocean \cite{stevens2020ocean}.
There are various factors that could play a role, such as air bubbles and impurities in the ice \cite{wengrove2023melting}, or meltwater flushing \cite{notz2009desalination}.
For nearshore sea ice, apart from the fact that the growth and melting of ice change the local depth of the liquid layer, the liquid layer deepens from the continent to the open ocean, which leads to the spatial variation in the thermal forcing.
In view of these, it would be both interesting and significant to further investigate the flow details inside mushy ice and bulk salty water through flow visualization and numerical simulation.
Future work could also explore ice evolution under more complex flow conditions and environmental settings, such as in the presence of a shearing mean flow, under periodic boundary temperatures, or with air bubbles and solid particles trapped in the ice.

The present study still leaves many open questions to be answered.
For example, it is worthwhile to combine experiments and numerical simulations to determine the heat and mass properties of the present system.
This will save the simplifications and empirical fittings in the parameterization, helping to establish a more accurate model of the system dynamics.
During the experiments, fluctuations of the ice front morphology are frequently observed (see supplementary videos), especially, there are sudden and rapid freezing and melting at the ice front, as reflected by the drastic changes of the ice-water lines in the space-time diagrams (Fig.~\ref{fig2}).
It would be very fruitful to have an in-depth examination of this phenomenon with detailed experimental measurements or numerical simulations on the ice structure and the flow field.
It would be also interesting to study the ice evolution on much longer time scales to check whether there exists any slower evolution beyond the final states of the experiments, or these states are the asymptotic steady state of the system.

\section{Acknowledgements}
We thank Rui Yang for insightful discussions. This work is supported by NSFC Excellence Research Group Program for ‘Multiscale Problems in Nonlinear Mechanics’ (No. 12588201), the Natural Science Foundation of China under Grant No. 12402321, the New Cornerstone Science Foundation through the New Cornerstone Investigator Program and the XPLORER PRIZE, the Postdoctoral Fellowship Program of the China Postdoctoral Science Foundation under Grant Nos. 2025T180524, GZB20240366 and 2024M751637, and Shuimu Tsinghua Scholar Program under Grant No. 2023SM038..

\section{Data avaliability}
All data supporting the findings of this study are included in the manuscript.

\section{Appendix A: Apparent velocity for salt diffusion}

The magnitude of the apparent velocity for vertical salt diffusion $v_D$ can be estimated by equating diffusive transport with convective transport as \cite{notz2009desalination, hoekstra1965migration}:
\begin{equation}
    \frac{S_c+S_{bulk}}{2}v_D=D_{brine}\frac{S_c-S_{bulk}}{h},
\label{eq:velocity}
\end{equation}
where $S_c$ is the local brine salinity at the cold top plate determined by $T_0(S_c)=T_c$, which is about $16.1\%$ in our experiments \cite{hall1988freezing}.
$S_{bulk}$ is the average salinity of the bulk salty water, which also approximates the local salinity at the ice front.
$D_{brine}$ is the solutal diffusivity of the brine in the mushy ice.
$h$ is the mean ice thickness from experiments.
We neglect the potential non-uniform distribution of local porosity and assume that that salinity profile in ice is approximately linear (quasi-steady diffusive or weak convective transfer in porous media).
With these assumptions, $(S_c+S_{bulk})/2$ on the left-hand side of Eq. \ref{eq:velocity} estimates the mean brine salinity in mushy ice, while the right-hand side estimates the salt flux.

\section{Appendix B: Model of heat and mass transfer}

For the heat transfer in the long term, as mentioned above, the heat fluxes on both sides of the ice front approximately equal after the initial rapid ice growth.
They can be estimated using the Fourier's law, with the enhancement of heat transfer due to convection in mushy ice and bulk liquid properly considered:
\begin{equation}
F=Nu_m\Lambda_m\frac{T_0(S_{bulk})-T_c}{h}=Nu_l\Lambda_l\frac{T_h-T_0(S_{bulk})}{H-h},
\label{eq:heat}
\end{equation}
where $F$ denotes the heat flux in the system, $Nu_m$ and $Nu_l$ are the Nusselt numbers in the mushy ice and bulk liquid that quantify the dimensionless heat transfer efficiency.
$\Lambda_m$ and $\Lambda_l$ are the thermal conductivities in the mushy ice layer and in the bulk liquid.
$\Lambda_m$ can be estimated as a linear combination of the thermal conductivities of brine and ice, $\Lambda_m=\phi\Lambda_{brine}+(1-\phi)\Lambda_{ice}$ where $\phi$ is the mean ice porosity.
$h$ and $H$ are the mean ice thickness and the system height.
$T_c$ and $T_h$ are the temperatures of the cold and hot plate.
$T_0(S_{bulk})$ is the freezing/melting point at the average bulk salinity, which is also approximately the temperature at the ice front.

For mass transfer on the scale of the entire ice layer, salt is gradually expelled from the ice layer into the bulk liquid.
As the ice thickness changes very slowly, this desalination can be estimated as:
\begin{equation}
        \frac{d}{dt}\left( \frac{S_c+S_{bulk}}{2} \phi\right)h=- Sh_mC\phi D_{brine} \frac{S_c-S_{bulk}}{h},
    \label{eq:salt}
\end{equation}
where $S_c$ is the local brine salinity at the cold top plate temperature determined by $T_0(S_c)=T_c$.
$S_{bulk}$ is the average salinity of the bulk salty water, which also approximates the local salinity at the ice front.
$h$ is the mean ice thickness.
$Sh_m$ is the Sherwood number in the mushy ice which quantifies the enhancement of dimensionless mass transfer efficiency due to convection.
$\phi$ is the mean porosity of the ice layer.
$D_{brine}$ is the solutal diffusivity of the brine in the mushy ice.
Noticing that Eq.~\ref{eq:velocity} underestimates the salt diffusion compared to the experiment observations, the mass transfer intensity can be corrected with the empirical parameter $C$.
Interestingly, in Fig.~\ref{fig2} and \ref{fig3}, the pore migration is observed to accelerate approaching the ice front.
While further investigations are need to obtain a better measurement, reveal the physical mechanism and make a theoretical prediction, this phenomenon hints that the desalination may be faster than that estimated with the mean salinity gradient, which also supports the need of this parameter.

Eq.~\ref{eq:salt} describes the relationship between the rate of change in the salt content in mushy ice and the salt flux through the ice front.
In addition to this equation, the salt content in mushy ice and the salt content in bulk liquid are also related by the mass conservation of salt:
\begin{equation} 
(S_{bulk}\rho_{bulk})|_{t=0}=\frac{S_c+S_{bulk}}{2}\rho_{brine}\cdot \phi\frac{h}{H}+S_{bulk}\rho_{bulk}\cdot (1-\frac{h}{H}),
\label{eq:mass}
\end{equation}
where $\rho_{bulk}$ and $\rho_{brine}$ are the average salty water density of the bulk liquid and the brine in mushy ice.
The volume change in the expansion vessel is neglected as it is much smaller than the volume of the cell.

As discussed in Appendix A, in Eqs.~\ref{eq:salt}-\ref{eq:mass}, we neglect the potential non-uniform distribution of local porosity and assume that that salinity profile in ice is approximately linear (quasi-steady diffusive or weak convective transfer in porous media).
With these assumptions, the mean brine salinity in ice can be estimated as $(S_c+S_{bulk})/2$.
Similarly, the mean temperature in ice can be estimated as $(T_c+T_0(S_{bulk}))/2$.
The average brine density $\rho_{brine}$ can then be estimated with these two estimations on the mean salinity and temperature.

In Eqs.~\ref{eq:heat}-\ref{eq:mass}, $T_c$, $T_h$, $H$ and $S_{bulk}|_{t=0}$ are the initial and boundary conditions of the experiments.
$\Lambda_m$, $\Lambda_l$, $T_0(S_{bulk})$, $S_c$, $D_{brine}$, $\rho_{bulk}|_{t=0}$, $\rho_{brine}$ and $\rho_{bulk}$ are the physical properties of ice and salty water that can be parameterized with these conditions and with $S_{bulk}$.
$C$ is an empirical coefficient.
Hence, $h(t)$, $S_{bulk}(t)$ and $\phi(t)$ can be solved numerically by inputting an initial condition, provided the parameterization of $Nu_m$, $Nu_l$ and $Sh_m$.

$Nu_m$ and $Sh_m$ quantify the enhanced heat/mass transfer efficiency relative to diffusion due to buoyant convection inside the mushy ice.
The strength of this convection is characterized by the Rayleigh number of the porous medium, $Ra_m$:
\begin{equation}
    Ra_m=\frac{Kgh}{\kappa_{brine}\nu_{brine}}\frac{\rho_c-\rho_0}{\rho_0},
    \label{eq:Ram}
\end{equation}
where $K$ is the permeability, $g=9.8$ m/s$^2$ is the gravitational acceleration, $h$ is the ice thickness, $\kappa_{brine}$ is the thermal diffusivity of brine in mushy ice, $\nu_{brine}$ is the kinetic viscosity of brine in mushy ice, $\rho_c$ is the brine density at the cold top plate and $\rho_0$ is the brine density at the ice front.

The permeability $K$ depends on the detailed ice structure and is positively related to ice porosity.
However, laboratory and field measurements show that it can vary by orders for different ice with the same porosity \cite{kawamura2006measurements, petrich2006modelling}.
Here $K$ (in m$^2$) is estimated with:
\begin{equation}
	K=\left\{
	\begin{aligned}
		&0 ,& \phi \le 0.054, \\
		&3\times10^{-11}(\phi-0.054)^{1.2} ,& \phi > 0.054,
	\end{aligned}
	\right .
        \label{eq:permeability}
\end{equation}
where $\phi$ is the average ice porosity.
The form of Eq.~\ref{eq:permeability} is consistent with that reported by Ref.~\cite{petrich2006modelling} and the estimated permeability is in good agreement with previous measurements.

As shown in Eq. \ref{eq:Ram}, the buoyancy that drives convection in mushy ice is induced by the difference in the density of the brine across the ice layer.
According to the equation of state of salty water \cite{gebhart1977new}, this density difference is mainly due to the salinity difference.
Ref.~\cite{bejan1985heat} shows that the mass transfer of convection driven by the salinity difference is comparable to the heat transfer of convection driven by the temperature difference.
Based on the experiment data by Ref.~\cite{gupta1973bounds}, $Sh_m$ in this work can be estimated with:
\begin{equation}
    Sh_m=1.3338+0.0099Ra_mLe_m, \ \  Ra_mLe_m>4\pi^2\approx39.5,
    \label{eq:Shm1}
\end{equation}
where $Le_m=\kappa_{brine}/D_{brine}$ is the Lewis number that represents the ratio between thermal diffusivity and solutal diffusivity in brine.

In the standard Rayleigh-B\'enard setting in an isotropic porous medium, the steady-state system is diffusive when the buoyancy intensity is lower than the critical value.
However, imposing $Sh_m=1$ for $Ra_mLe_m\le4\pi^2$ will lead to a discontinuity in $Sh_m$ according to the fitting results (Eq.~\ref{eq:Shm1}).
Meanwhile, the mushy ice layer in the present study is not exactly the same as the typical settings.
For example, the ice layer exchanges heat and mass at its bottom with the bulk salty water and it can have a complex porous structure.
The buoyancy intensity changes continuously with time.
Besides, the permeability is difficult to be accurately parameterized when the porosity is low, which affects the estimation of the buoyancy intensity.
These also bring challenges to parameterize the transport properties under low buoyancy conditions, making it difficult to determine the threshold for convection in our experiments.
For practical reasons, in this work, a linear interpolation between $Sh_m|_{Ra_mLe_m=0}=1$ and $Sh_m|_{Ra_mLe_m=4\pi^2}=1.725$ is adopted:
\begin{equation}
    Sh_m=1+0.0184Ra_mLe_m, \ \  Ra_mLe_m\le4\pi^2\approx39.5.
        \label{eq:Shm2}
\end{equation}

In the present work, the buoyancy intensity is rather limited ($Ra_mLe_m|_{max}\approx100$).
It is expected that brine convection is still weak and diffusion should still play a crucial role in the heat and mass transfer.
Therefore, $Nu_m$ should not deviate much from $Sh_m$, unlike the case in high buoyancy regime.
They are jointly estimated as:
\begin{equation}
	Nu_m\approx Sh_m=\left\{
	\begin{aligned}
		&1+0.0184Ra_mLe_m,& Ra_mLe_m\le4\pi^2\approx39.5, \\
		&1.3338+0.0099Ra_mLe_m,& Ra_mLe_m>4\pi^2\approx39.5.
	\end{aligned}
	\right .
        \label{eq:NumShm}
\end{equation}

For $Nu_l$, it quantifies the enhancement of heat transfer by convection in the bulk liquid.
It depends on the Rayleigh number $Ra_l$:
\begin{equation}
    Ra_l=\frac{g(H-h)^3}{\kappa_l\nu_l}\frac{\rho_0-\rho_h}{\rho_h},
    \label{eq:Ral}
\end{equation}
where $g$ is the gravitational acceleration, $H$ is the height of the system, $h$ is the ice thickness, $\kappa_l$ is the thermal diffusivity of the bulk liquid, $\nu_l$ is the kinetic viscosity of bulk liquid, $\rho_0$ is the brine density at the ice front and $\rho_h$ is the brine density at the hot bottom plate.

In the $Ra_l$ regime of this work ($10^7<Ra_l<10^9$), the simulation results of Ref.~\cite{wang2021growth} suggest that $Nu_l$ in the fresh water below a growing ice layer follows the scaling of $(Ra_l-Ra_{lcr})^{0.27}$ where $Ra_{lcr}=1708$.
However, the aspect ratio of the system can affect the convection structure.
Freezing and melting in salty water may also involve double-diffusive effects \cite{wettlaufer1997natural, guo2025effects}.
Therefore, the pre-factor is selected as 0.23 by matching the asymptotic final ice thickness in both cases:
\begin{equation}
    Nu_l=0.23(Ra_l-Ra_{lcr})^{0.27}.
        \label{eq:Nul}
\end{equation}

With Eqs. \ref{eq:Ram}-\ref{eq:Nul}, $Nu_m$, $Sh_m$, $Nu_l$ and the related physical properties can be parameterized with $h$, $S_{bulk}$, $\phi$, as well as the initial and boundary conditions of the experiments.

In this work, we choose the empirical parameter $C=1.5$ and inputting $\phi=0.08$ at $t=1.5$ day as the parameterization and initial condition to best fit the experiment results.

\section{Appendix C: Freezing/melting point of NaCl solution as function of salinity}
\begin{figure*}[!htb]
\centering
\includegraphics[width=0.4\linewidth]{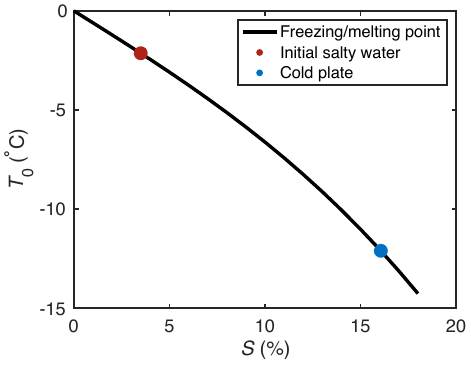}
\caption{The freezing/melting point of NaCl solution as function of the salinity \cite{hall1988freezing}.
The red point denotes the initial salty water in this study, with $S=3.5\%$ and $T_0(S)=-2.1$ $^\circ$C.
The blue point denotes the condition at the cold top plate, with $S_c=16.1\%$ and $T_0(S_c)=T_c=-12.1$ $^\circ$C.
}
\label{fig8}
\end{figure*}

\section{Appendix D: List of supplemental material movies}

\noindent Movie S1: Ice evolution in the experiment of $\Theta_i=0.44$, played at 86400$\times$ playback speed.\\
\noindent Movie S2: Ice evolution in the experiment of $\Theta_i=0.29$, played at 86400$\times$ playback speed.

\normalem

\end{document}